\documentstyle[preprint,aps]{revtex}

\begin{document}
\draft
\title{Spontaneous Oscillations of Collective Molecular Motors}
\author{Frank J\"ulicher and Jacques Prost}
\date{\today}
\address{Institut Curie, Physicochimie Curie, 
Section de Recherche, 26 rue d'Ulm,
75231 Paris Cedex 05, France}
\maketitle

\begin{abstract} \hskip 0.15in
We analyze a simple stochastic model to describe motor molecules which
cooperate in large groups and present a physical mechanism 
which can lead to oscillatory motion if the motors are
elastically coupled to their environment. Beyond a critical fuel 
concentration, the non-moving state of the system becomes unstable
with respect to a mode with angular frequency $\omega$. We present
a perturbative description of the system near the instability
and demonstrate that oscillation frequencies are determined by the
typical timescales of the motors.
\end{abstract}

Motor proteins are highly specialized macromolecules which can consume
mechanical energy to induce motion and to generate forces. 
These molecules are involved in active transport processes,
cell locomotion and muscle contraction \cite{albe94,huxl69}.
A typical motor molecule specifically attaches to a certain protein
filament which serves as a track for its motion. In the presence
of fuel, which in the cell ist adenosinetriphosphate (ATP), the
motor starts moving in a direction defined by the polarity of the
track filament.
Experimental methods to study the physical properties
of motor proteins have been developed in
recent years \cite{spud90,ishi91,svob93,hunt94,wink95bour96}. 
These methods allow to measure forces and velocities of
individual motors or small groups of motors. 

Normally, at given fuel concentration and temperature, motor molecules
will generate a constant average force leading to a constant average
velocity \cite{wink95bour96}. 
In some cases, however, biological
motors are used to drive oscillatory motion.
Insect flight muscles e.g. generate oscillating forces in order to
move the wings with high frequency. While for some insects
(e.g. butterflies) the timing
of these oscillations is controlled by a periodic external 
nerve signal, others (e.g. bees and wasps) generate oscillations
within the muscle. 
The mechanisms which lead to these oscillations are not understood
but they seem to be related to the process of force-generation 
\cite{prin78}.
 
The purpose of this article is to describe theoretically 
a physical mechanism by which motor collections
acting on a spring can give rise to oscillatory motion.
Oscillations occur as the result of
cooperation of many motors while individual
motors only generate time-independent velocities.

The microscopic mechanisms which are responsible for the force generation
of molecular motors are not understood. In order to indentify possible 
physical mechanisms
which allow such a motor molecule to
convert the chemical energy of a fuel to motion
and mechanical work, simple physical models have been suggested
\cite{ajda92,pros94chau94,pesk94,astu94,magn9394,doer95,dere96,zhou96}.
One idea \cite{pros94chau94} is to simplify the internal degrees
of freedom of the motor to two different states of a particles
which moves along a one-dimensional coordinate $x$.
The interaction of the particle with the track is described by
periodic potentials which reflect the periodic structure 
along the surface of the track.
Within such a two-state model
motion is induced if the potentials 
are asymmetric with respect to $x\rightarrow -x$ 
(which reflects the polarity of the track)
and if detailed balance of the transitions between states
is broken \cite{pros94chau94}. 

New phenomena occur if the collective motion of many 
particles is considered \cite{leib93,juli95,dere9596}. 
In a simple model for collective
motion, particles 
are rigidly coupled to a common
structure which they set in motion collectively \cite{juli95}. 
As a result of cooperation, 
motion can in this case even 
occur in a symmetric system via spontaneous symmetry 
breaking. 
Instabilities and dynamical phase transitions also
occur in asymmetric systems revealing the rich 
behaviour of many-motor systems. At these transitions, a steady state
becomes unstable and the system chooses a new
steady state. These transitions correspond therefore to dynamical 
instabilities at angular frequency $\omega=0$.

In the following, we demonstrate that many coupled motors also
can induce instabilities with $\omega\neq 0$ towards a state
with periodically varying velocity. Such instabilities occur if the system
is elastically coupled to its environment. This
implies, that the complete system is connected 
to its environment via
a spring attached to the common backbone, see Fig. \ref{f:sys} \cite{fn2}. 
A single particle reaches in this case a steady state
with zero velocity and a position at which the average force generated
by the motor is balanced by the elastic force of the spring \cite{svob93}.

We are interested in the behavior of many motors coupled together.
Let us first review the most simple two-state model which we are 
going to discuss \cite{pros94chau94,juli95}.
Motors are described by particles which move
in two periodic potentials $W_\sigma(x)$ with period $l$.
Transitions between states $\sigma=1,2$
occur with rates $\omega_{1}(x)$ and $\omega_{2}(x)$ which also
are $l$-periodic. 
The excitation amplitude can in general be written as \cite{pros94chau94}
$\omega_1(x)=\omega_2(x)[\exp[(W_1-W_2)/T] + \Omega\theta(x)]$,
where $T$ denotes temperature measured in units of the Boltzmann constant.
We assume that 
the periodic function $\theta(x)$ is given. In principle it can be 
calculated from a specific model for the reaction kinetics of
the fuel coupled to the particle \cite{juli96}. 
The amplitude $\Omega$ is related to the 
difference of chemical potentials
$\Delta \mu \equiv \mu_{ATP} - \mu_{ADP}-\mu_{P}$
of the fuel and its products which is the chemical driving force of the
motors. Here, we assume that the fuel is ATP which is hydrolyzed
to adenosinediphosphate (ADP) and phosphate (P):
$ATP\rightleftharpoons ADP+P$.
For $\Delta\mu=0$, $\Omega=0$ and the rates
$\omega_{1}$ and $\omega_{2}$ obey detailed balance
$\omega_{1}/\omega_{2}=exp[(W_1-W_2)/T]$.
As soon as $\Delta \mu \neq 0$,
$\Omega$ becomes nonzero.
For small $\Delta\mu/T$, $\Omega\sim\Delta\mu$; for large $\Delta\mu/T$,
$\Omega\sim C_{ATP}$ is proportional to the fuel concentration.

We now study the behavior of many particles which are rigidly
connected to a common backbone, see Fig. \ref{f:sys}. 
Particles have a fixed spacing $q$ which for simplicity is 
assumed to be incommensurate with the period $l$,
i.e. $l/q$ is irrational. In the limit of an infinite system one
can then introduce densities $P_\sigma(\xi)$ with $\sigma=1$,  $2$
which give the probability to find a particle at position
$\xi= x \; \hbox{\rm mod}\;  l$ relative to the beginning of the potential
period in state $\sigma$. These densities are not independent but
obey $P_1(\xi)+P_2(\xi)=1/l$ and $\int_0^l d\xi (P_1+P_2)=1$.
The equations of motion for this system read \cite{juli95}
\begin{eqnarray}
\partial_t P_1   &+& v\partial_\xi P_1  = - \omega_{1} P_1 +\omega_{2} P_2 
\nonumber\\
\partial_t P_2   &+& v\partial_\xi P_2  =   \omega_{1} P_1 -\omega_{2} P_2
 \label{eq:P12} \\
f_{ext} & = &\lambda v + KX +\int_0^l  d\xi P_1 \partial_{\xi}(
W_1 - W_2 )\label{eq:fext} \quad .
\end{eqnarray}
Eq. (\ref{eq:P12}) 
describes the dynamics of the density $P_1(\xi)$
resultig from motion of the backbone with velocity $v=\partial_t X$ and the
transitions between the states. Eq. (\ref{eq:fext}) is a force
balance: The externally applied force per particle $f_{ext}(t)$
is balanced by viscous drag with damping coefficient $\lambda$,
the average force excerted by the potentials and an additional elastic
force $KX$ of a spring of length $X$ 
with elastic modulus $K N$ where $N$ is the number
of particles. Note, that Eqns. (\ref{eq:P12}) and
(\ref{eq:fext}) only depend on the difference $W_1-W_2$ 
of the two potentials and we can choose $W_2$ to
be constant without loss of generality.

As a consequence of the spring action, a nonmoving
solution to Eqns. (\ref{eq:P12}) and (\ref{eq:fext}) exists for $f_{ext}=0$,
$v=0$, $P_1=R\equiv\omega_{2}/(\alpha l)$, 
where 
$\alpha(\xi)\equiv\omega_{1}(\xi)+\omega_{2}(\xi)$
and 
$X=X_0\equiv -\int_0^l d\xi R\partial_\xi(W_1-W_2)/K$.
We first study
the linear stability of this solution and determine the instability threshold
with respect to oscillations. With the ansatz
$P_1(\xi,t)  = R(\xi)+ p(\xi) \exp(s t)$, $v(t)  = u \exp(st)$
and $X(t) =  X_0 + u \exp(st)/s$,
one finds using Eq. (\ref{eq:P12}) and (\ref{eq:fext})
to linear order in $p$ and the velocity
amplitude $u$:
\begin{equation}
p(\xi) = -u \frac{\partial_\xi R}
{s+\alpha(\xi)} \quad .
\end{equation}
The possible values of the complex eigenvalue $s=-\tau+i\omega$ 
are determined by
\begin{equation}
\lambda + \frac{K}{s} = \int_0^l d\xi \frac{\partial_\xi R\; 
\partial_\xi (W_1-W_2)}
{s+\alpha(\xi)} \quad .\label{eq:complexs}
\end{equation}
The non-moving state is unstable if $\tau <0$. The instability occurs
for $\tau=0$, where the real part of $s$ vanishes. This happens at
a threshold value $\Omega=\Omega_c(K)$ for which $\alpha(\xi)=\alpha_c(\xi)$.
This critical value together with the frequency $\omega_c$ 
at the instability
is determined  by 
\begin{eqnarray}
\lambda & = &\int_0^l d\xi \frac{\alpha_c}
{\alpha_c^2 +\omega_c^2} \partial_\xi R \;\partial_\xi (W_1-W_2)
\label{eq:lambda}\\
 K & = & \int_0^l d\xi 
\frac{\omega_c^2}{\alpha_c^2+\omega_c^2}
\partial_\xi R \; \partial_\xi(W_1-W_2)
\quad \label{eq:kl}.
\end{eqnarray}
In the limit $K=0$ where no elastic element is present,
it follows from Eq. (\ref{eq:kl}) that $\omega_c=0$. Eq. (\ref{eq:lambda})
now determines $\Omega_c(0)$ which is the condition for the instability
of the non-moving state as previously described in Ref. \cite{juli95}. 
In the case of a symmetric system, this instability
leads to spontaneous motion via a symmetry-breaking transition.
For nonzero $K$ the instability occurs 
for $\Omega_c(K)=\Omega_c(0)+\delta\Omega_c$ with
$\delta\Omega_c\sim K$ for small $K$ and the system starts
to oscillate with finite angular frequency
$\omega_c \sim K^{1/2}$. Note, that beyond a maximal value $K>K_{max}$,
the resting state is stable for any value of $\Omega$.

Fig. \ref{f:vt} (a) and (b) show 
examples for these oscillations which have been
calculated numerically
for constant deexcitation rate $\omega_{2}$, $\lambda \omega_2 l^2/U=0.1$
and piecewise linear and symmetric  potentials
$W_\sigma(\xi)$, see Fig. \ref{f:sys} (a) with $a=l/2$. In this
case $K_{max}l^2/U\simeq 2.8$.
The function $\theta(\xi)$ is chosen as
shown in Fig. \ref{f:sys} (a) with $d/l=0.1$.
Fig. \ref{f:vt} (a)
displays the position $X$ versus time $t$ for
an excitation level $\Omega=0.1$ slightly above threshold and elastic modulus
$Kl^2/U=0.2$.
After an initial relaxation period, motion is
almost sinusoidal. An example for small elastic molulus $Kl^2/U=0.002$
and same $\Omega$
is shown in Fig. \ref{f:vt} (b). The periodic motion in this
case shows cusp-like maxima which correspond to sudden
changes of the velocity. These cusps are related to discontinuities 
of the velocity as a function of external force for $K=0$ \cite{juli95}.

We now give an analytical description of oscillations in
the vicinity of the instability. 
Anticipating, that the motion is periodic with
period $t_P\equiv 2\pi/\omega$, we can write
$P_1(\xi,t) = \sum_{k=-\infty}^\infty P_1(\xi,k) e^{ik\omega  t}$,
$v(t) = \sum_{k=-\infty}^\infty v_k e^{ik\omega t}$, and
$f_{ext}(t) = \sum_{k=-\infty}^\infty f_k e^{ik\omega  t}$,
which defines the Fourier coefficients $P_1(\xi,k)$, $v_k$ and
$f_k$. Using this representation, one can derive the nonlinear
relation between velocity and external force
\begin{equation}
f_k =
F^{(1)}_{kl} v_l + F^{(2)}_{klm} v_l v_m + F^{(3)}_{klmn}v_l v_m v_n 
+ O(v^4) \quad .\label{eq:fnvn}
\end{equation}
The coefficients $F^{(n)}_{k,k_1,..,k_n}$ can be calculated
by first rewriting
Eq. (\ref{eq:P12}) as
\begin{equation}
P_1(\xi,k) = \delta_{k,0} R(\xi) 
-\sum_{lm}\frac{\delta_{k,l+m}}{\alpha + i\omega k}v_l \partial_\xi
P_1(\xi,m) \quad .\label{eq:P1f}
\end{equation}
Inserting the ansatz
\begin{equation}
P_1(\xi,k) = R\delta_{k,0} + P^{(1)}_{kl}(\xi) v_l + P^{(2)}_{klm}
(\xi) v_l v_m + O(v^3) 
\quad ,
\end{equation}
into Eq. (\ref{eq:P1f}), one obtains a recursion relation
for the functions $P^{(n)}_{k, k_1,..,k_n}$:
\begin{equation}
P^{(n)}(\xi)_{k, k_1, .. ,k_n}= -\sum_{l}\frac{\delta_{k,k_n+l}}
{\alpha + i\omega k} \partial_\xi P^{(n-1)}_{l, k_1, ..,k_{n-1}} \label{eq:rec}
\end{equation}
Using Eq. (\ref{eq:fext}), one finds
\begin{equation}
F^{(1)}_{kl} \equiv \delta_{kl}\left(
 \lambda + \frac{K}{i\omega k} - \int_0^l d\xi
\frac{\partial_\xi R\partial_\xi(W_1-W_2)}{\alpha + i\omega k}\right)
\quad .
\end{equation}
The linear response function of the
system is given by the inverse of $F^{(1)}_{kl}$. 
Therefore the instability condition
Eq. (\ref{eq:complexs}) corresponds to $F^{(1)}_{kl}=0$.
For $n>1$,
\begin{equation}
F^{(n)}_{k,k_1,..,k_n}\equiv \int_0^l d\xi P^{(n)}(\xi)_{k,k_1,..,k_n}
\partial_\xi(W_1-W_2) \quad .\label{eq:Fn}
\end{equation}
Note, that the coefficients $F^{(n)}_{k,k_1,..,k_n}$ are nonzero
only if $k=k_1+..+k_n$. This results from time translational
symmetry \cite{fn1}. 
If no elastic coupling to the environment is present, i.e. 
$K=0$, one recovers for constant external force $f_0$
the steady states which have been described
previously \cite{juli95}: 
$f_{ext}=f_0(v_0)$; $v_n=0$ for $n\neq 0$. As soon
as $K\neq 0$, $v_0$ must vanish and a constant external force $f_0$
only changes the average 
position $X_0$.

Spontaneous oscillations are solutions to
Eq. (\ref{eq:fnvn}) for $f_n=0$. The
dominant terms near the instability of $v_1$
are given by
\begin{eqnarray}
0 & = & F^{(1)}_{11} v_1 + G^{(2)} v_{-1} v_2 + G^{(3)} v_1^2 v_{-1}
\label{eq:v1} \\
0 & = & F^{(1)}_{22} v_2 + F^{(2)}_{211} v_1^2  \quad ,\label{eq:v2}
\end{eqnarray}
where $G^{(2)}\equiv F^{(2)}_{1,2,-1}+F^{(2)}_{1,-1,2}$ and
$G^{(3)}\equiv F^{(3)}_{1, 1,1,-1} + F^{(3)}_{1, 1,-1,1} +
F^{(3)}_{1, -1,1,1} $. As soon as $v_1$ and $v_2$ are determined, higher
orders $v_n$ can be obtained recursively using
Eq, ({\ref{eq:fnvn}).
From Eq. (\ref{eq:v2}) one finds that
$v_2 \sim  v_1^2$.
Inserting this value in Eq. (\ref{eq:v1}), one obtains
\begin{equation}
0 = F^{(1)}_{11}v_1 + \tilde G^{(3)} v_1^2 v_{-1} \label{eq:v1r} \quad ,
\end{equation}
with an effective third order coefficient
$\tilde G^{(3)}\equiv
G^{(3)}-F^{(2)}_{211}G^{(2)}/F^{(1)}_{22}$.
One solution to Eq. (\ref{eq:v1r}) is always $v_1=0$.
The remaining 
solutions are described by
\begin{equation}
|v_1|^2 = -F^{(1)}_{11}/\tilde G^{(3)} \quad .\label{eq:v1v1}
\end{equation}
Since the amplitude 
$|v_1|^2$ is a real number, solutions to Eq. (\ref{eq:v1v1})
exist only if the right hand side of Eq. 
(\ref{eq:v1v1}) is real and positive. In general, however, $F^{(1)}_{11}$
and $\tilde G^{(3)}$ are complex numbers. The 
requirement to find real solutions to Eq. (\ref{eq:v1v1})
fixes the oscillation frequency. This can be seen as follows:
If the system is at its instability threshold $\Omega=\Omega_c(K)$, 
$F^{(1)}_{11}=0$ for $\omega=\omega_c$, 
see Eq. (\ref{eq:complexs}).
In the $(\Omega,\omega)$-plane, a line exists along which
$F^{(1)}_{11}/\tilde G^{(3)}$ is real. This line can be parametrized
by a function $\omega=\omega_s(\Omega)$ which
at the instability threshold obeys
$\omega_c=\omega_s(\Omega_c)$. For $\Omega>\Omega_c$,
$-F^{(1)}_{11}/\tilde G^{(3)}$ is real and positive along this line thus 
allowing for a
solution to Eq. (\ref{eq:v1v1}).
For $\Omega<\Omega_c$, $v_1=0$ is the only solution.
Fig. \ref{f:vt} (c) shows
the dependence of the selected frequency
$\omega_s$ as a function of the excitation amplitude.
For $\Omega<\Omega_c(K)$, $\omega_s=0$. At the instability, a finite frequency 
is selected which increases for increasing $\Omega$.

Note, that this example is developed for a symmetric system for which 
an additional restriction holds:
Eq. (\ref{eq:fnvn}) is now invariant with respect to the transformation
$v_n\rightarrow -v_n$, $f_n\rightarrow -f_n$.
As a result, all even expansion coefficients in Eq. (\ref{eq:fnvn})
vanish, 
$F^{(2)}=F^{(4)}=..=0$. In particular, $G^{(2)}$ vanishes
and $\tilde G^{(3)}=G^{(3)}$. 
Furthermore, all terms which couple $v_1$ to modes $v_n$ with even $n$ vanish
and $v_2=v_4=..=0$. 
As a result,
the periodic motion of a symmetric system has the symmetry property
$v(t+t_P/2)=-v(t)$. 

The parameters used in Fig. \ref{f:vt} correspond e.g. to the
choice $\lambda \simeq 10^{-8} kg/s$, $\omega_2\simeq 10^{3} s^{-1}$, 
$l\simeq 10^{-8} m$,
$U\simeq 10 k_B T$. Oscillation frequencies in this case 
vary between $\omega_s=0$ for $K=0$
and $\omega_s \simeq 5 \omega_2$ for
$K \simeq K_{max} \simeq 10^{-3}N/m$ per motor \cite{fn3}. 
This shows that using typical time scales of biological motors
the mechanism described here generates
frequencies up to the $kHz$ range.

In summary, we have shown using a simple two-state model for
foce-generation of molecular motors that oscillatory motion can occur
as a collective effect. Motors which individually
only generate constant forces, show oscillating behavior if they
are coupled in large numbers and work against an elastic spring.
Beyond a critical amplitude of excitations which drive the system, 
a nonmoving state becomes unstable with respect to periodic motion.
This argument suggests that oscillations are a possible
mode of motion of biological motor proteins and gives an example of how
some insect flight muscles could generate oscillations via the
mechanism of force generation.

We thank F. Gittes, A. Ott, L. Peliti, D. Riveline and 
W. Zimmermann for
useful discussions and A. Ajdari for a critical reading of the
manuscript. We are also very grateful to
F. Amblard, M. Bornens, J. K\"as and D. Louvard for information
concerning spontaneous oscillations in biological systems. 
F.J. acknowledges a Marie Curie Fellowship of the
European Community; part of this work was done at the Aspen center
for physics.

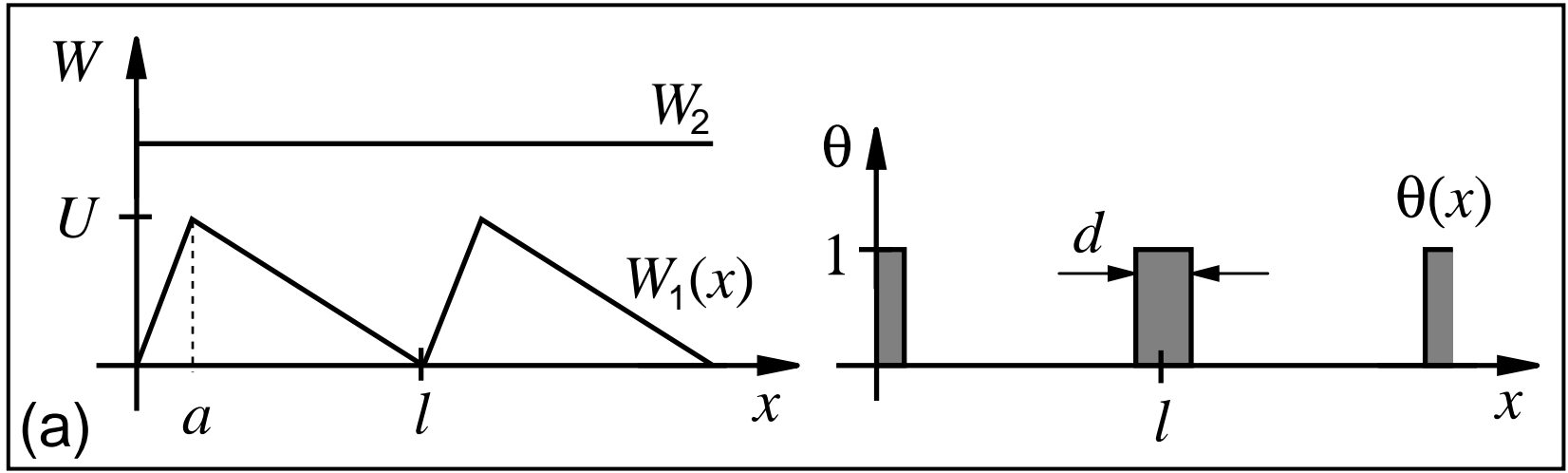
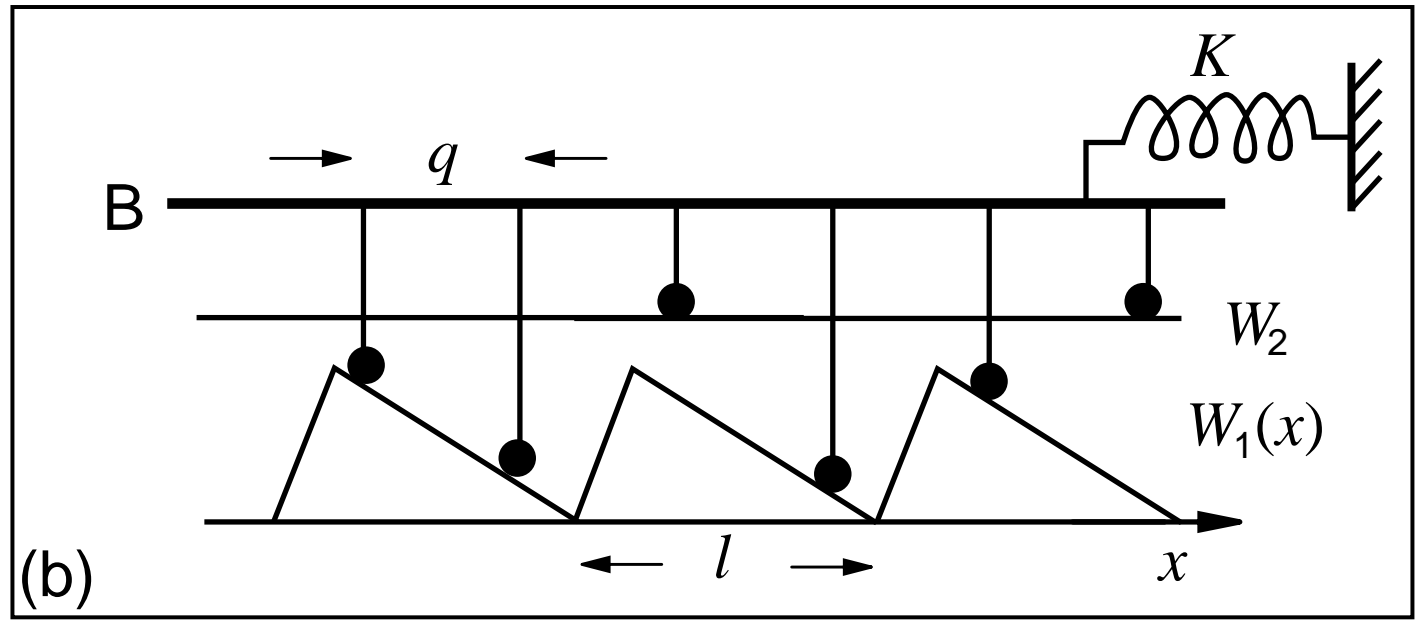
\begin{figure}
\caption{(a) Periodic potentials $W_1$ and 
$W_2$ used to calculate the behavior of the system.
The Function $\theta(x)$
is chosen to be nonzero in the vicinity of the potential minimum.
(b) Many particles coupled rigidly to a backbone which is connected 
with its environment via a spring K. The particle
spacing is denoted $q$.}
\label{f:sys}
\end{figure}

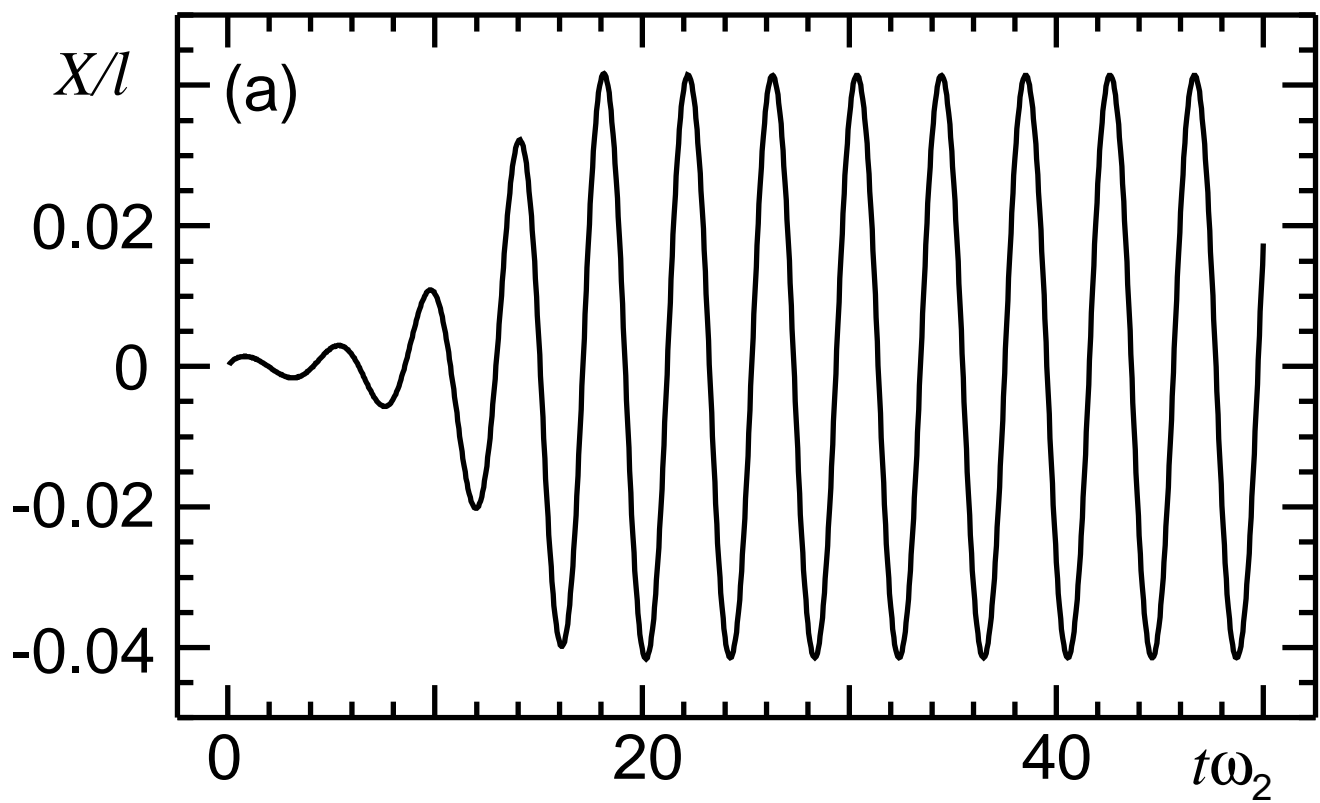
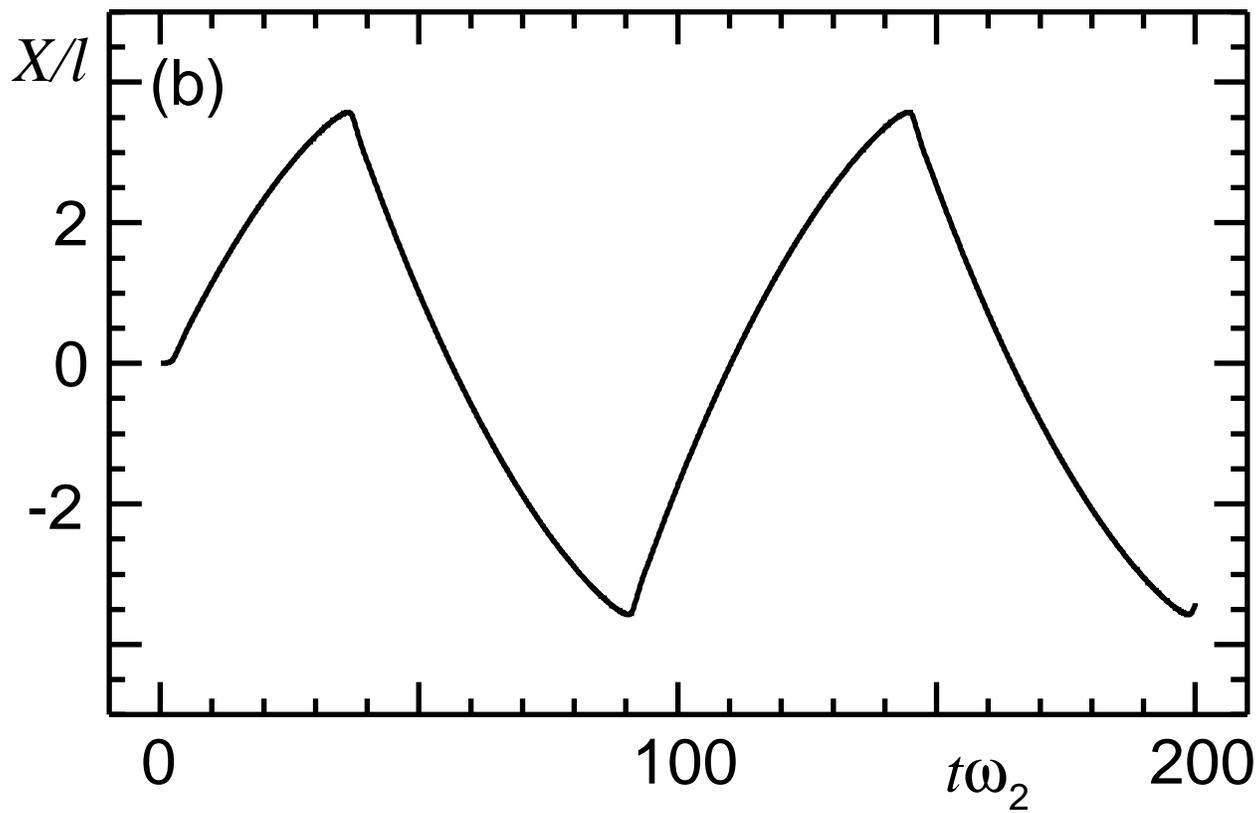
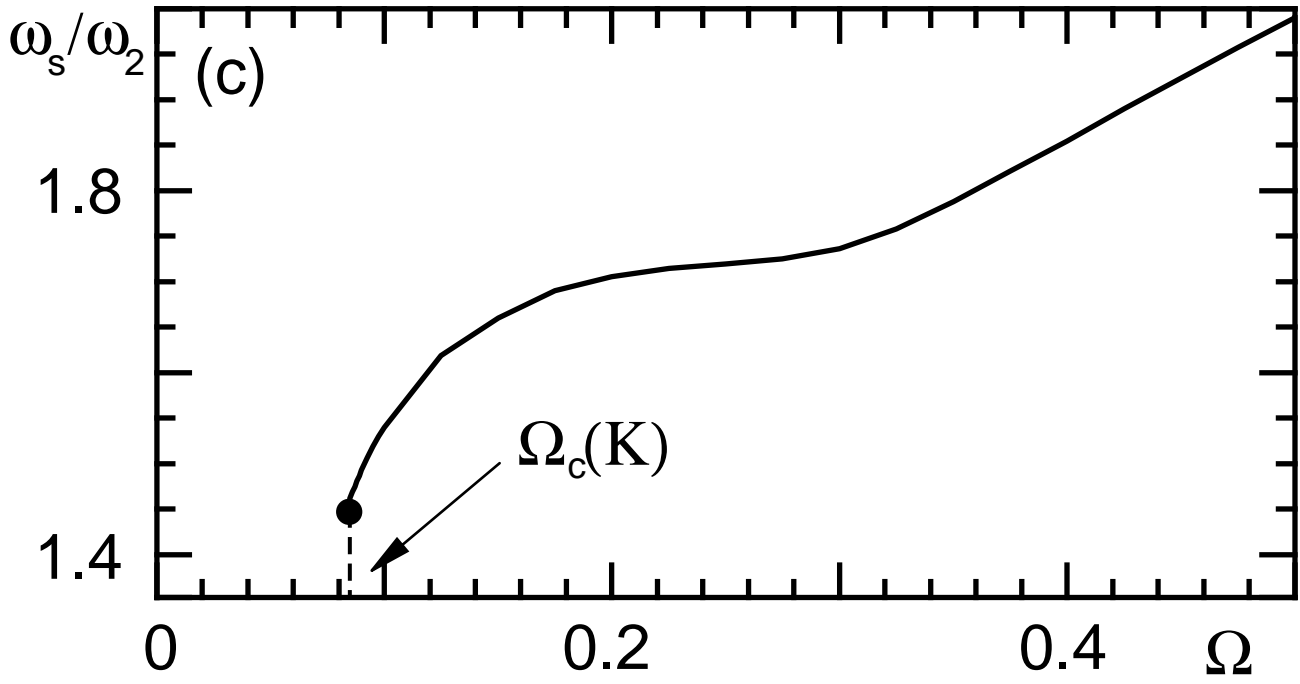
\begin{figure}
\caption{(a) Position $X$ versus
time $t$ for a symmetric system with $d/l=0.1$,
$\Omega=0.1$, $\lambda\omega_2 l^2/U=0.1$
and $K l^2/U=0.2$ (see Fig. \protect\ref{f:sys} (a)). 
(b) Position versus time
for same system but $K l^2/U=0.002$. (c) 
Selected frequency $\omega_s$ as a function of the excitation
amplitude $\Omega$ for the case $\Omega=0.1$, $K l^2/U=0.2$. The critical
excitation amplitude is $\Omega_c\simeq 0.08$. }
\label{f:vt}
\end{figure}


\begin{references}
\bibitem{fn3} Here, we have assumed the limit of large $\Omega$
corresponding to saturated ATP concentration.
For a system of $10^3$ motors, $K_{max}$ corresponds to an
elastic modulus of $1 N/m$.
\end{references}
\end{document}